\documentstyle[12pt,twoside]{article}
\begin{document}
\begin{center}

{\Large\bf SINGULARITY-FREE\\[5PT]
COSMOLOGICAL SOLUTIONS IN\\[5PT]
STRING THEORIES\\[5pt]}
\medskip
 
{\bf
J. C. Fabris\footnote{e-mail: fabris@cce.ufes.br} and R. G. Furtado\footnote{e-mail: furtado@cce.ufes.br}} \medskip

Departamento de F\'{\i}sica, Universidade Federal do Esp\'{\i}rito Santo, 
29060-900, Vit\'oria, Esp\'{\i}rito Santo, Brazil
\medskip

\end{center}
 
\begin{abstract}
Singularity-free cosmological solutions may be obtained from the string action
at tree level if the dimension of the space-time is greater than $10$ and if
brane configurations are taken into account. The behaviour
of the dilaton field in this case is also regular. Asymptotically a radiative phase is
attained indicating a smooth transition to the standard cosmological model. 
\vspace{0.7cm}

PACS number(s): 04.20.Cv., 04.20.Me
\end{abstract}

At tree level, the string action can be reduced to the gravitational
action coupled non-minimally to the dilaton field and to gauge fields.
These gauge fields may couple non-minimally or minimally to the dilaton
field depending if they come from the Neveu-Schwartz or Ramond-Ramond sector
of the string action \cite{lidsey}. In general, the dimension of the space-time in string
theories is equal to $10$ if supersymmetry is considered. Bosonic string may
leave in a $26$ dimensional space-time. But, $M$-theory requires a $11$-dimensional
space-time while the $F$-theory is generally formulated in $12$ dimensions \cite{polchinski,green,kiritsis}.
\par
Cosmology is an arena where the consequences of string theories can be explored,
since the typical string effects must manifestate themselves at very large energy levels,
which
can be attained only in the very early universe. One of the expectation concerning
string cosmology relies on the possibility to obtain singularity-free primordial
cosmological models. However, this expectation has been frustrated until now: even
when a non-singular four dimensional space-time is obtained, divergences in the
dilaton field appear and employement of an effective action at tree level becomes
doubtfull \cite{picco,kirill}. In some cases, as the singularity is approached non-linear geometrical terms
may be taken into account, avoiding the appearence of divergences in the curvature
invariants. But, the entire scenario is composed of many branchs rendering all the
model quite artificial \cite{gasperini}.
\par
In this work we consider the string effective action in four dimension obtained from
the original $D$-dimensional theory by dimensional reduction and truncation. This effective
action will be coupled to ordinary radiative matter, which may be considered as a
manifestation of the electromagnetic term existing in the Neveu-Schwartz and/or
Ramond-Ramond sector. Singularity-free solutions are found. But, in order to obtain
a complete regular model, in the geometric and dilaton fields, the dimension of the
space-time must be greater than $10$. The recent activity with the so-called $M$-theory and $F$-theory, as well as with another string-type models in dimensions greater than $10$, may
render the model developped here attractive. One important ingredient to obtain the
regular models found here is to consider $p$-branes configuration. In particular,
eleven and twelve dimensional space-times with a $4$-brane are quite interesting models, representing
some of the cases where a complete regular scenario is obtained.
\par
Let us consider the string effective action at tree level:
\begin{equation}
L = \sqrt{-\tilde g}e^{-\tilde\sigma}\biggr\{\tilde R  - \tilde\omega\tilde\sigma_{;A}\tilde\sigma^{;A} -
\frac{1}{12}H_{ABC}H^{ABC}\biggl\}
\end{equation}
where $\tilde\sigma$ is the dilatonic field, $H_{ABC}$ is the axionic field and
$\tilde\omega$ is the dilatonic coupling constant. The tildes indicate that all
quantities are considered in a $D$-dimensional space-time.
In particular, the dilatonic coupling constant, when a $p$-brane configuration is
taken into account is given by \cite{duff}
\begin{equation}
\tilde\omega = - \biggr\{\frac{(D-1)(p-1) - (p+1)^2}{(D-2)(p-1) - (p+1)^2}\biggl\}
\end{equation}
where $p$ denotes the order of the brane: $p = 0$ indicates a pointlike configuration,
$p = 1$ a stringlike configuration, $p = 2$ a membrane, and so on. When $p = 1$, $\tilde\omega = - 1$ for any dimension of the space-time.
\par
The $D$-dimensional metric is written as
\begin{equation}
ds^2 = g_{\mu\nu}dx^\mu dx^\nu - e^{2\beta}\delta_{ij}dx^idx^j \quad,
\end{equation}
where $g_{\mu\nu}$ is the four dimensional metric, $e^\beta$ is the scale factor
of the $d$-dimensional internal space which we suppose to be homogenous and flat.
Hence we obtain the following effective action in four dimension
\begin{equation}
\label{l1}
{\it L} = \sqrt{-g}\phi\biggr[R - \gamma_{;\rho}\gamma^{;\rho} -
\omega\frac{\phi_{;\rho}\phi^{;\rho}}{\phi^2} - 
\phi^{n-1}\Psi_{;\rho}\Psi^{;\rho}\biggl] + {\it L_m}.
\end{equation}
In this action,
\begin{eqnarray}
\phi = e^{d\beta-\tilde\sigma} \quad &,& \quad \gamma = a\beta+b\ln\phi \quad , \\
a = \{d(d+1)+\tilde\omega d^2\}^{1/2} \quad &,& \quad b = - d(1+\tilde\omega)\{d(d+1)+\tilde\omega d^2\}^{-1/2} \nonumber
\quad .
\end{eqnarray}
The field $\Psi$ comes from the axionic term.  The parameter
$n$ was included in order to consider other effective actions, for example, those
coming from supergravities or pure multidimensional space-time \cite{picco}. The string case with
an axionic field corresponds to $n = - 1$.
The term ${\it L_m}$ represents the ordinary matter. Here, we will be
interested in the radiative fluid only, since 
it seems more realistic when we have in mind the primordial Universe. Moreover,
a radiative fluid may be obtained through the reduction of string action to
four dimensions. For example, the 5-form existing in the Ramond-Ramond sector of
the superstring type IIB leads in four dimension to, besides other scalar fields,
an eletromagnetic field without coupling with the dilaton or moduli fields. This
is a general feature of $D/2$-forms in a $D$-dimensional space-time \cite{fabris}. The new coupling
constant is given by
\begin{equation}
\omega = - \biggr\{\frac{(d - 1)\tilde\omega + d}{(d + 1)\tilde\omega + 1}\biggl\}
\quad .
\end{equation}
Notice that, when $\tilde\omega = - 1$, $\omega = - 1$. Hence, the pure string case
is a fixed point.
\par
From (\ref{l1}) we obtain the field equations:
\begin{eqnarray}
R_{\mu\nu} - \frac{1}{2}g_{\mu\nu}R &=& 8\pi\frac{T}{\phi} + \frac{\omega}{\phi^2}\biggr(\phi_{;\mu}\phi_{;\nu} - \frac{1}{2}g_{\mu\nu}\phi_{;\rho}\phi^{\rho}\biggl) + \frac{1}{\phi}\biggr(\phi_{\mu\nu}
- g_{\mu\nu}\Box\phi\biggl) + \nonumber\\
&+& \biggr(\gamma_{;\mu}\gamma_{;\nu} - \frac{1}{2}g_{\mu\nu}\gamma_{;\rho}\gamma^{;\rho}\biggl) + \phi^{n - 1}\biggr(\Psi_{;\mu}
\Psi_{;\nu} - \frac{1}{2}g_{\mu\nu}\Psi_{;\rho}\Psi^{;\rho}\biggl) \quad ;\\
\Box\phi + \frac{1 - n}{3 + 2\omega}\phi^n\Psi_{;\rho}\Psi^{;\rho} &=& \frac{8\pi}{3 + 2\omega}T \quad ;\\
\Box\Psi + n\Psi_{\rho}\frac{\phi^{;\rho}}{\phi} &=& 0 \quad ; \\
\Box\gamma + \gamma_{;\rho}\frac{\phi^{;\rho}}{\phi} &=& 0 \quad ; \\
{T^{\mu\nu}}_{;\mu} &=& 0 \quad .
\end{eqnarray}
Considering the Friedmann-Robertson-Walker metric
\begin{equation}
ds^2 = dt^2 - a(t)^2\biggr(\frac{dr^2}{1 - kr^2} + r^2(d\theta^2 + \sin^2\theta d\phi^2)\biggl) \quad ,
\end{equation}
$k$ being the curvature of the spatial section ($k = 0, 1, -1$ for a flat, close and
opened model, respectively),
the field equations reduce to the following equations of motion:
\begin{eqnarray}
\label{em1}
3\biggr(\frac{\dot a}{a}\biggl)^2 + 3\frac{k}{a^2} &=& 8\pi\frac{\rho}{\phi} +
\frac{\omega}{2}\biggr(\frac{\dot\phi}{\phi}\biggl)^2 - 3\frac{\dot a}{a}\frac{\dot\phi}{\phi} + \frac{\dot\gamma^2}{2} + \phi^{n-1}\frac{\dot\Psi^2}{2} \quad ;
\\
\label{em2}
\ddot\phi + 3\frac{\dot a}{a}\dot\phi + \frac{1 - n}{3 + 2\omega}\phi^n\dot\Psi^2 &=&
\frac{8\pi}{3 + 2\omega}(\rho - 3p) \quad ; \\
\label{em3}
\ddot\Psi + 3 \frac{\dot a}{a}\dot\Psi + n\dot\Psi\frac{\dot\phi}{\phi} &=& 0 \quad ;\\
\label{em4}
\ddot\gamma + 3\frac{\dot a}{a}\dot\gamma + \dot\gamma\frac{\dot\phi}{\phi} &=& 0 \quad ;\\
\label{em5}
\dot\rho + 3\frac{\dot a}{a}(\rho + p) &=& 0 \quad .
\end{eqnarray}
In these expression, $\rho$ is the density of ordinary matter and
$p$ is the pressure which obeys a barotropic equation of state, $p = \alpha\rho$.
\par
The equations (\ref{em3},\ref{em4},\ref{em5}) admit the first integrals:
\begin{equation}
\dot\Psi = \frac{A}{a^3\phi^n} \quad , \quad \dot\gamma = \frac{B}{a^3\phi} \quad ,
\quad \rho = \rho_0a^{3(1 + \alpha)} \quad ,
\end{equation}
where $A$, $B$ and $\rho_0$ are integration constants.
Let us now specialize the equations for the radiative fluid case ($\alpha = 1/3$).
Equation (\ref{em2}) becomes
\begin{equation}
\ddot\phi + 3\frac{\dot a}{a}\dot\phi + \frac{1 - n}{3 + 2\omega}\frac{A^2}{a^6\phi^n} = 0
\end{equation}
which can be reduced to the integral
\begin{equation}
\label{int}
\int\frac{d\phi}{\sqrt{1 - \frac{2A^2}{(3 + 2\omega)C}\phi^{1 - n}}} = \sqrt{C}\theta \quad 
\end{equation}
where $C$ is another integration constant, and $\theta$ is a new time parameter such
that $dt = a^3d\theta$.
\par
This integral can be explicitly solved. The case $\gamma =$ constant and $\omega > - 3/2$
has been studied in \cite{picco}. Bouncing solutions, with no curvature singularity,
were obtained when $- 3/2 < \omega < - 4/3$. In one asymptotic the dilaton field takes
a constant value; but in the other asymptotic it diverges, leading to a divergence in
the string coupling parameter, $g_s = \phi^{-1}$. This fact may render the effective action
senseless. Perhaps this problem may be coped with through the duality properties of
string theory. But, it is not sure that such duality properties can be applied to the
background defined by the solution found in \cite{picco}.
Here we exploit the possibility that $\omega < - 3/2$. In this case the integral (\ref{int})
can be solved through the redefinition of time parameter. The solution for $\phi$ takes
the form,
\begin{equation}
\phi = \phi_0(\sinh\xi)^{2/(1 - n)} \quad , \quad \phi_0 = \biggr\{\frac{(3 + 2\omega)C}{2A^2}\biggl\}^{1/(1 - n)} \quad .
\end{equation}
In this expression, $\xi$ is the time parameter connected with the cosmic time by
\begin{equation}
dt = \frac{2}{(1 - n)\sqrt{C}}\phi_0 a^3(\sinh\xi)^{(1 + n)/(1 - n)} d\xi \quad .
\end{equation}
\par
In order to solve (\ref{em1}) and determine the behaviour of the scale factor,
we write $a = \phi^{-1/2}b$, rewrite the resulting expression in terms of the
time parameter $\xi$, obtaining the relation
\begin{equation}
\frac{b'}{b} = \pm \sqrt{Mb^2 - kb^4 + B^2 - 1}\frac{\sqrt{|3 + 2\omega|}}{1 - n}\frac{1}{\sinh\xi}
\end{equation}
where the prime means derivative with respect to $\xi$.
Let us consider the case where $k = 0$ and $B = 0$. This implies a flat universe where,
in principle, the internal scale factor is constant. However, notice the it can
correspond to a time-dependent internal scale factor which bears a specific relation
with the field $\phi$. In any case, the scale factor of the external space takes the
form
\begin{equation}
\label{sol}
a = - a_0[(\sinh\xi)^{-1/(1 - n)}]\times\frac{1}{\cos[\ln(\tanh^r\xi/2)]} \quad , \quad r =
\pm\frac{\sqrt{|3 + 2\omega|}}{1 - n} \quad .
\end{equation}
This solution represents a bouncing universe in the interval
\begin{eqnarray}
\xi_i < \xi < x_f \quad , \quad \xi_{i,f} = \ln\biggr[\frac{1 + s_{i,f}}{1 - s_{i,f}}\biggl] \quad , \nonumber \\
 \quad s_{i,f} = e^{(n_{i,f} + 1/2)\frac{\pi}{r}} \quad ,
\quad n_i = n_f - 1 \quad , \quad n_f < 0 \quad .
\end{eqnarray}
The cosmic time, in this interval, is such that $- \infty < t < \infty$. This assures
the absence of any curvature singularity. Moreover, the field $\phi$ takes finite
non-zero
values during all the evolution of the universe. Hence, there is no divergence of the
string coupling parameter, even in the extreme of the interval. There is also no
divergence in the scale factor (what is trivial if we consider the constant value,
of course). So, the solution described by (\ref{sol}) is a complete regular solution.
In both asymptotic the scale factor behaves as $a \propto t^{1/2}$ what allows for
a smooth transition to a radiative phase and consequently to the standard cosmological
model.
\par
The most important difference beteewn the results found here and those presented in \cite{kirill} is
the absence of a divergence in the string perturbative coupling parameter in the
initial asymptotic. This is due to the presence of a radiative fluid, a case not
considered in that work. It must also be stressed that in
\cite{kirill} the dynamics of the internal space was considered in its full complexity,
while here just some particular cases were exploited. But, as it has been already remarked in \cite{kirill},
the possibility to obtain singularity-free solutions arises only if the dimension
of the space-time is greater than $10$, since anomolous theories are obtained in
the Einstein-frame formulation of the effective action.
\par
In fact, the question which arises now concerns if this singularity-free solution can be
implemented in the context of string theories. One of the essential features of the
model exposed above is
that $\omega < - 3/2$. This can be only achieved if the dimension of the space-time
is such that $D > 10$ for some brane configurations. It excludes the traditional superstring
case. However, one example where it can be obtained is in the realm
of the so-called $M$-theory and $F$-theory, which are formulated in $D = 11$ and
$D = 12$ respectively. Both the $M$-theory and $F$-theory are connected
with string theories, since the superstring models may be viewed as specific
background configurations of those higher dimensional theories \cite{polchinski,clifford}. For example, for the case $B = 0$, with $D = 11$, $D = 12$ with $p = 4$  we obtain
$\omega = \tilde\omega = - 5/2$ and $- 8/5$, respectively. These are interesting situations because one can relate them
with the brane world program \cite{langlois} by which we live in a four-dimensional brane. It must also
be stressed that the regular solutions displayed here are valid for $n = - 1$, which is
the typical value of this parameter for string theories.
Notice, however, that in the present letter we have exploited only one of the simplest case.
But, our goal here is to show that a complete singularity-free solution, even in the
dilatonic field, may be obtained
in the context of the string action if we allow the dimension of the space-time be greater
than 10. A more complete analysis may reveal other possible situations. Finally, it must be remarked that the gravitational coupling is
connected with the inverse of the field $\phi$, which takes constant values in both
asymptotics in such a way that the value of the gravitational coupling in the first asymptotic is greater than its value in the second asymptotic, what opens the possibility
to solve the hierarchial problem of the cosmological constant in a way similar to the
the brane cosmology program. In the case treated here, the gravitational coupling dicreases its value by a factor of $10^6$ between the initial and final asymptotics
when $\omega = - 8/6$ ($D = 12$, $p = 4$).
\newline
\vspace{1.0cm}
{\bf Acknowledgement:} We thank CNPq (Brazil) for partial financial support.

\end{document}